\documentstyle[11pt,newpasp,twoside,epsf]{article} 
\def\edcomment#1{\iffalse\marginpar{\raggedright\sl#1\/}\else\relax\fi}
\reversemarginpar

\begin{document}
\title{Does TV Col Have the longest Recorded Positive Superhumps?}
\author{Alon Retter, Coel Hellier}
\affil{Physics Dept., Keele University, Staffordshire, ST5 5BG, UK}
\author{Tomas Augusteijn}
\affil{Isaac Newton Group of Telescopes, Apartado 321, 38700 Santa Cruz 
de La Palma, Canary Islands, Spain}
\author{Tim Naylor}
\affil{School of Physics, University of Exeter, Stocker Road, Exeter,\\ 
EX4 4QL, UK}

\begin{abstract}
Re-examination of extensive photometric data of TV~Col reveals evidence 
for a permanent positive superhump. Its period (6.4~h) is 16 percent 
longer than the orbital period and obeys the well known relation between 
superhump period excess and binary period. At 5.5-h, TV~Col has an 
orbital period longer than any known superhumping cataclysmic variable 
and, therefore, a mass ratio which might be outside the range at which 
superhumps can occur according to the current theory. We suggest several 
solutions for this problem. 
\end{abstract}

\section{Introduction}
\subsection{Permanent superhumps}

Permanent superhump systems compose a new subclass of cataclysmic variables 
(CVs), whose existence was established only in the nineties. Systems that 
belong to this group have quasi-periodicities slightly shifted from their 
orbital periods, in addition to the binary periods themselves. Unlike 
SU~UMa systems (see Warner 1995 for a review on SU~UMa systems and CVs in 
general), which show this behaviour only during superoutbursts, in permanent 
superhump systems the phenomenon is observed during their normal brightness 
state. According to Osaki (1996), permanent superhumpers differ from other 
subclasses of non-magnetic CVs by their relatively short orbital periods 
and high mass transfer rates, resulting in accretion discs that are 
thermally stable but tidally unstable. Retter \& Naylor (2000) provided 
observational support for this idea.

The `$\bf positive$ $\bf superhump$', periodicity which is a few percent 
larger than the orbital period, is explained as the beat period between 
the binary motion and the precession of an eccentric disc in the apsidal 
plane. Periods slightly shorter than the orbital periods have also been 
seen in several systems. They are known as `$\bf negative$ $\bf superhumps$' 
(Patterson 1999). The observations show a roughly linear relation between 
the positive superhump period excess as a fraction of the binary period 
and the  binary period (Stolz \& Schoembs 1984). Negative superhumps seem 
to obey a similar rule (Patterson 1999). It has been suggested that negative 
superhumps are generated by the nodal precession of the accretion disc 
(Patterson et al. 1993; Patterson 1999); however, there are some theoretical 
difficulties with this idea (e.g. Murray \& Armitage 1998).

\subsection{Periodicities in TV~Col}

The periodicities detected so far in TV~Col and their common interpretations
are (Motch 1981; Hutchings et al. 1981; Schrijver, Brinkman \& van der Woerd 
1987; Barrett, O'Donoghue \& Warner 1988; Hellier, Mason \& Mittaz 1991; 
Hellier 1993; Augusteijn et al. 1994):

\begin{itemize}

\item 4 day - the nodal precession of the accretion disc

\item 5.5 hr - the orbital period

\item 5.2 hr - the negative superhump (the beat between the orbital period 
and the nodal precession)

\item 32 min - the spin period

\end{itemize}

There seems to be a strong connection between positive and negative 
superhumps. Light curves of many permanent superhumpers show both types 
of superhumps. In addition, period deficits in negative superhumps are 
about half period excesses in positive superhumps (Patterson 1999): 
$\epsilon_{negative}$$\approx$--0.5$\epsilon_{positive}$, where 
$\epsilon$=($P_{superhump}$--$P_{orbital})/P_{orbital}$. 
We, therefore, decided to look in available photometric data on TV~Col
for positive superhumps, which would be predicted to have a period near 
6.4 h. Here we report the finding of such a periodicity as initially 
announced by Retter \& Hellier (2000). 

\section{Observations}

We have reanalysed existing optical photometry that was presented by 
Barrett et al. (1988); Hellier et al. (1991); Hellier (1993); Hellier \& 
Buckley (1993) and Augusteijn et al. (1994). This dataset contains 77 
nights of observations obtained during 1985-1991 from eight separate runs. 
The telescopes used were the South African Astronomical Observatory (SAAO) 
0.75-m \& 1-m telescopes and the Dutch 0.91-m telescope at the European 
Southern Observatory (ESO). Runs with mini-outbursts (see e.g. Augusteijn 
et al. 1994) were rejected for obvious reasons. The 1989 January run is the 
best among the six remaining datasets. It consists of six successive nights, 
each longer than six hours; no outburst occurred during the run; the 
observations were carried out with a CCD (rather than a photometer) and 
with the largest telescope among the three used, accumulated more photons 
(`clear' filter), and had the longest exposure times (less dead-time), thus 
giving the best signal to noise ratio among all datasets. 


\section{Analysis}

In Fig.~1a the power spectrum of the data from the best run is shown. The 
detrending was done by subtracting the mean from each night. In addition 
to the two known periods (5.5 h and 5.2 h, marked as f1 and f2 
respectively) and their aliases, there is a third peak (labelled f3) 
and its aliases. These peaks also appear in the raw power spectrum when 
no detrending method is used. After fitting and subtracting the two known
frequencies (f1 and f2), f3 and its 1-d alias become the two strongest 
peaks in the residual power spectrum (Fig.~1b). We chose f3 as the real 
period rather than its 1-d alias (which is stronger) for reasons given 
below (see Fig.~2). Furthermore, the same relative peak strengths is seen 
in the orbital period, where the alias is stronger than the true period.

\begin{figure}

\centerline{\epsfxsize=3.0in\epsfbox{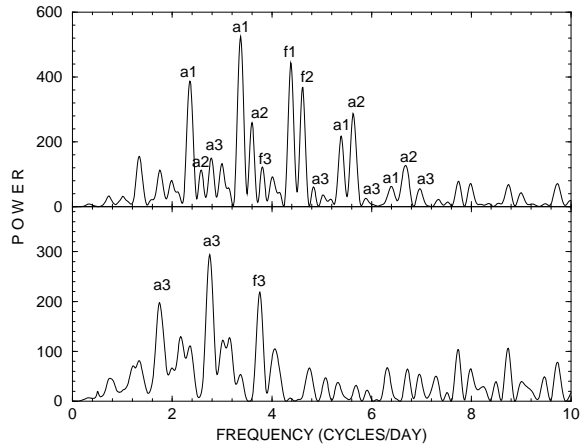}}

\caption{Power spectra of the 1989 January run. 
Upper panel (a): In addition to the two previously known periods -- the 
orbital period (f1) and the negative superhump (f2), there is a third 
structure of peaks centered at a 1-d alias of 6.4~h (f3); `ai' (i=1,2,3) 
represent aliases of `fi' correspondingly;
Lower panel (b): After subtracting f1 and f2, the candidate period is still 
present.}


\end{figure}


We ran a few tests to check the significance of the candidate period,
0.265(3) d. Here we give details of what we consider the most important 
test. We tried to assess the probability that correlated noise could be 
responsible for the candidate periodicity. In the absence of a model for 
the correlated noise, the best test is to use the repeatability in 
different datasets. Given that we found a period in the best set, we can 
ask how likely it is that the strongest period in the other datasets 
(after the known periods had been subtracted) would be consistent with it. 
The probability of the highest peak in another dataset being, by chance, 
compatible with the candidate period in the best set is 0.08. This was 
calculated from (i) the frequency error for the candidate period, which 
implies that the peaks are identical if they are within 0.04 day$^{-1}$, 
and (ii) the range over which it could occur taken as the spacing of the 1 
day aliases (1 day$^{-1}$ is the maximum range over which periods are truly 
independent). The period discovered in the best set was seen in two of the 
remaining five datasets. We thus used the binomial distribution to find that 
the probability of this occurring by chance was formally 5 percent. However, 
two of the sets that do not show the period are those with the shortest 
runs each night, and thus the lowest data quality. Thus the 95\%  
significance value should be regarded as a lower limit.

\section{Discussion}

The photometric data show evidence for a periodicity of 0.265 d in 
addition to the previously known periods. The repeatability of the peak 
in three independent datasets makes it 95\% significant. In addition, the 
better the data are, the more the period stands out of the noise. Moreover, 
it has almost exactly the value predicted from the Stolz \& Schoembs (1984) 
relation (updated by Patterson 1999) shown in Fig.~2. TV~Col has already 
been classified as a permanent superhump system because its 5.2-h period 
was interpreted as a negative superhump. In addition, the new period and 
the negative superhump obey the relation between the two types of 
superhumps (Section~1.2). Therefore, the new period is naturally interpreted 
as a positive superhump. 

\begin{figure}

\centerline{\epsfxsize=2.75in\epsfbox{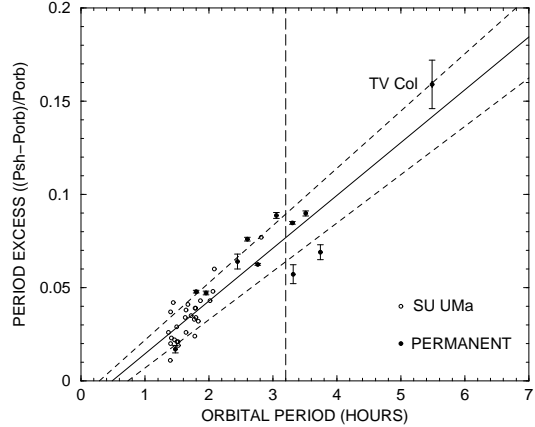}}

\caption{The relation between superhump period excess (over the binary
period) and binary period in superhump systems. Empty circles correspond 
to periods in the SU~UMa systems. Filled circles represent permanent 
superhumpers. The solid line represents the linear fit to the data. The 
two tilted dashed lines show the 1-$\sigma$ error. TV~Col obeys the 
relation. The upper edge of the period gap (as defined by Diaz \& Bruch 
1997) is marked by the vertical long-dashed line.}



\end{figure}

According to theory superhumps can appear only in CVs with small mass 
ratios -- q=$ M_{donor}/M_{compact}$$<$0.33. Hellier (1993) concluded, 
however, that q=0.62-0.93 from a spectroscopic analysis of the system, but 
this depended on an interpretation of the emission lines that may not be 
correct. Using the superhump excess, we find: q=0.95$\bf \pm$0.10 -- well 
above the 0.33 limit suggested by the hydrodynamic simulations, and 
consistent with the values estimated by Hellier. The mass ratio in TV~Col
may thus be above the theoretical limit, perhaps due to its strong 
magnetic field. Alternatively, TV~Col may be an extreme system, with a very 
massive white dwarf near the Chandrasekhar mass (1.44$\bf M_{\odot}$), and 
/ or an undermassive secondary star.

\section{Summary}

\begin{itemize}

\item Analysis of previously published photometric data reveals 
evidence for the presence of an additional period (6.4~h) in the optical 
light curve of TV~Col. This periodicity can be identified with the 
positive superhump.

\item Our findings support the classification of TV~Col as a permanent 
superhump system. TV~Col has, therefore, the longest known superhump 
period among all CVs.

\item TV~Col offers a unique opportunity to test and reject some of the 
models as it extends the superhump regime to periods far beyond the predicted 
values, where the difference between the models become significant. The 
mass ratio of TV~Col might exceed the limit for superhump systems allowed 
by hydrodynamic simulations. Therefore, a confirmation for our findings is 
urgently required by further observations.

\end{itemize}

\end{document}